\documentclass[a4paper,11pt]{article}
\usepackage[utf8]{inputenc}
\usepackage[T1]{fontenc}
\usepackage{lmodern}
\usepackage[zerostyle=e]{newtxtt}
\usepackage{newtxtext}
\usepackage{comment}
\usepackage{latexsym}
\usepackage{mathrsfs}
\usepackage{amsbsy}
\usepackage{amssymb}
\usepackage{amsfonts}
\usepackage{amsmath}
\usepackage{amsthm}
\usepackage{newtxmath}
\usepackage{enumitem}
\usepackage{xcolor}
\usepackage{color}
\usepackage{mathtools}
\usepackage{caption} 
\usepackage{subcaption} %sous-figures 
\usepackage{euscript} % FOR HAND CALLIGRAPHIC 
\usepackage{float}
\usepackage{diagbox}
\usepackage[pdftex]{graphicx}
\usepackage{expl3}
\usepackage{xparse}
\usepackage[font=small,labelfont=bf]{caption}
\usepackage[colorlinks=true
,urlcolor=posurl
,anchorcolor=posurl
,citecolor=posurl
,filecolor=posurl
,linkcolor=posurl
,menucolor=posurl
,linktocpage=true
,pdfa=true
]{hyperref}
\usepackage[title]{appendix}
\usepackage{tabularx}
\usepackage[numbers,sort&compress]{natbib}
\usepackage{pos}

\title{A formalism for the ambiguities of the Wheeler-DeWitt equation}
\author*[a]{Eftychios Kaimakkamis}
\author[a,b]{Karunava Sil}
\affiliation[a]{Department of Physics, University of Cyprus,\\
  Nicosia 1678, Cyprus}

\affiliation[b]{Department of Physics, New Alipore College,\\
L Block, New Alipore, Kolkata 700053, India}

\emailAdd{ekaima02@ucy.ac.cy}
\emailAdd{karunavasil@gmail.com}
\abstract{We study ambiguities in the precise formulation of the Wheeler-DeWitt equation for the wavefunction of the Universe that arise due to different operator orderings in the quantum Hamiltonian. We first examine the simpler case of the 1-dimensional minisuperspace model and derive the inner product measure that renders the Hamiltonian hermitian. Based on this, we establish that the Hilbert space inner products and quantum probabilities are universal, free of any ambiguities, at the semiclassical level. Recasting the Wheeler-DeWitt equation in a form invariant under field redefinitions of the minisuperspace variable, we show that all ambiguity functions are contained in a higher order scalar function, which can be used to define classes of models with universal predictions to all orders in $\hbar$. We then generalize to minisuperspace models of arbitrary dimension, upon the inclusion of an arbitrary number of scalar matter fields. We show that the hermiticity of the Hamiltonian in these cases provides a set of constraints, which can be used to cast the WDW equation in a covariant form and establish the universality of the inner products at the semiclassical level. In these cases as well, all ambiguity functions appear in a higher order scalar function of the minisuperspace manifold, which leads to distinct universality classes.}

\FullConference{Corfu Summer Institute 2023 "School and Workshops on Elementary Particle Physics and Gravity" (CORFU2023)\\
 23 April - 6 May , and 27 August - 1 October, 2023\\
Corfu, Greece\\}

\begin{document}
\maketitle

\section{Introduction}

\noindent There are a number of open questions regarding the Wheeler-DeWitt (WDW) equation for the wavefunction of the Universe \cite{dewitt, hawking, isham}. 
%is known for having a number of open questions and unresolved issues . 
Particularly important is the issue of ambiguities regarding operator orderings, which arise upon the quantization of the classical Hamiltonian. Different operator orderings lead to different WDW equations for the wavefunction of the Universe, denoted througout this article $\Psi$. The resulting ambiguities can be parameterized in terms of apriori undetermined functions in the differential equation \cite{toumbas}, with each choice leading to distinct solutions for the wavefunction. Naively, it seems that the same classical theory leads to different quantum prescriptions, yielding distinct results for the Hilbert space inner products and the quantum probabilities.
%solutions. Upon calculating different inner products, this in turn seems to be leading to conflicting results for the same measurable observables. \\

In this work, we first resolve the issue of operator ordering ambiguities at the semiclassical level for the simpler case of the one-dimensional minisuperspace model \cite{toumbas, kehagias}, where the only dynamical degree of freedom is the scale factor of the Universe. Specifically, we show that the inner products and probabilities associated with the various quantum prescriptions are universal at the semiclassical level \cite{toumbas, kehagias}. For this purpose, we use the inner product measure $\mu$ that renders the quantum Hamiltonian hermitian \cite{toumbas} to define the wavefunction $\widehat{\Psi}=\sqrt{\mu}\,\Psi$, which transforms as a scalar under field redefinitions of the scale factor. The square of the absolute value of $\widehat{\Psi}$ determines the invariant probability density to find the Universe in a certain quantum state. We obtain the WDW equation for $\widehat{\Psi}$ and cast it in an explicitly invariant form under field redefinitions. The universality at the semi-classical level is made manifest since in the equation for $\widehat \Psi$, all ambiguity functions associated with the various operator orderings appear in a higher order (in $\hbar$) correction term $\mathcal{Z}$, which is a scalar under field redefinitions. Because this term does not contribute at the semiclassical level, we obtain a universal expression for $\widehat{\Psi}$ and, hence, for the invariant quantum probability density. Demanding further that the inner products and, therefore, $\widehat{\Psi}$ are ambiguity-free to all orders in $\hbar$ fixes the scalar function $\mathcal{Z}$ to be independent of the ambiguity functions, and imposes certain relations between the ambiguity functions. In a forthcoming article \cite{paper} we compute $\mathcal{Z}$ for the path integral wavefunctions whose measures are based on field redefinitions of the scale factor, which were studied in \cite{toumbas} at the semiclassical level, in order to investigate whether these belong to the same universality class to all orders in $\hbar$. 

We next proceed to incorporate an arbitrary number of scalar matter fields $\phi^{\alpha}$, which we take to be minimally coupled to gravity. We allow for a non-trivial field configuration manifold and generic interactions among the scalar fields. In addition, we allow for anisotropies in the spacetime metric but retain the homogeneity property. 
%After some reformulations, 
In the minisuperspace approximation, we end up with a non-linear $\sigma$-model with a one-dimensional base and a non-trivial Lorentzian target space manifold. A much greater number of ambiguity functions appear now in the WDW equation due to the presence of a greater number of dynamical degrees of freedom. The hermiticity of the Hamiltonian does not uniquely fix the inner product measure as in the one-dimensional case but provides a number of constraints, which we derive. Defining $\widehat{\Psi}$ as before and imposing the hermiticity constraints, we obtain the WDW equation for $\widehat{\Psi}$ and establish universality at the semi-classical level. 
%In addition, we show that by finding solutions to this equation, the measure is unique chosen after all. 
All ambiguity functions appear in a higher order in $\hbar$ scalar function $\mathcal{Z}$. The universality of the inner products to all orders in $\hbar$ requires $\mathcal{Z}$ to be ambiguity-free and limits the possible values.

\section{1d minisuperspace model}

\noindent To start with consider a cosmological model in the minisuperspace approximation in the presence of a positive cosmological constant $\Lambda$ and in the absence of matter. The Lorentzian action is
\begin{equation}\label{action1}
S= \frac{1}{2}\int \limits_{\mathcal{V}} d^4x \sqrt{-g}\left(R-2\Lambda\right),
\end{equation}
where $\mathcal{M}$ denotes the spacetime manifold\footnote{
%Due to $\mathcal{M}$ in the minisuperspace approximation having no boundary $\partial \mathcal{M}$, 
When $\mathcal{V}$ has a boundary, we must add the Gibbons-York-Hawking boundary term: $-\int \limits_{\partial \mathcal{V}}d^3x \sqrt{h} \, K $ , where $h_{ij}$ is the induced metric on the boundary and $K$ is the extrinsic curvature \cite{wheeler}.}. As an ansatz, we insert a FRLW metric:
\begin{equation}
\label{FRW'}
ds^{2} = -N^{2}(t) dt^{2} + a^2(t)\left (\frac{dr^{2}}{1-kr^{2}} + r^{2} d\Omega_{3}^2 \right ).
\end{equation}
In this equation, $N(t)$ and $a(t)$ are the lapse function and the scale factor, respectively, while $d\Omega _{3}^{2}$ is the metric of a unit $3$-sphere. The lapse function $N$ is non-dynamical and can be fixed by time a reparametrization. The scale factor $a$ is the sole dynamical degree of freedom. The constant $k$ takes the values $-1$, $0$, or $1$ for an open, flat or closed Universe, respectively. For closed cosmologies, we set $k=1$ and the metric takes the form
\begin{equation}
\label{FRW}
ds^{2} = -N^{2}(t) dt^{2} + a^2(t)d\Omega_{3}^2.
\end{equation}
Substituting the metric above in the action, we obtain the effective Lagrangian for the scale factor
\begin{equation} \label{lagrangian}
L= 3v_{3}N \Biggl(-\frac{a\dot{a}^{2}}{N^2}+a-\ell ^{2}a^3\Biggr),
\end{equation}
where $v_{3} = 2\pi^{2}$ and $\ell = \sqrt{\frac{\Lambda}{3}}$. The model can be exhibited as a one-dimensional sigma model \cite{toumbas, dewitt, kiefer, page, linde}
\begin{equation} \label{minisuperspace}
    L = N \left [\gamma _{aa} \frac{\dot{a}^{2}}{N^{2}} - V(a) \right ],
\end{equation}
where $\gamma _{aa} = -3v_{3}a$ is the metric on the one-dimensional target space manifold, which is Lorentzian, and $V = -3v_{3}a\left (1- \ell ^{2} a^{2} \right )$ is the potential. By finding the conjugate momentum $\pi _{a} = \gamma _{aa}\frac{2\dot{a}}{N}$, it is then straightforward to obtain the classical Hamiltonian
\begin{equation} \label{hamiltonian}
    \frac{H}{N} = \frac{\partial L}{\partial N} = \left ( \frac{1}{4} \gamma ^{aa} \pi _{a}^{2} + V \right ) = 0.
\end{equation}
The fact this is vanishing can be seen to be a consequence of the time reparametrization symmetry of the theory. 

In order to quantize the Hamiltonian above, we promote the conjugate momentum of the scale factor to a quantum operator 
\begin{equation} \label{canmom1}
    \pi_{a} \rightarrow -i\hbar \frac{d}{da}
\end{equation}
giving the following canonical commutation relation
\begin{equation} \label{canonical1}
\begin{split}
    [\pi_{a},a] &= -i\hbar.
\end{split}
\end{equation}
The quantization gives rise to a problem regarding the precise ordering of the operators in the quantum Hamiltonian \cite{toumbas, dewitt, hawking, kiefer}. More specifically, while classically the kinetic component of the Hamiltonian may be rewritten with the momenta and the minisuperspace metric in any order we wish, the different reorderings are not equivalent at the quantum level. Generally, we can always introduce a pair of ambiguity functions \cite{toumbas}:
\begin{equation} \label{eq:ambiguities}
    \gamma ^{aa} \pi _{a} \pi _{a} = \frac{1}{\rho _{1} \rho _{2}} \gamma ^{aa} \pi_{a} \, \rho _{1} \, \pi _{a} \, \rho _{2}.
\end{equation}
This classically constitutes a tautology, but after quantization the expression accounts for a variety of distinct terms, given that the ambiguity functions depend on $a$ and are, therefore, affected by the derivatives. Notice that we do not allow $\rho _{1}$ and $\rho _{2}$ to be functions of $\pi _{a}$. We shall also restrict ourselves to the case of these functions being strictly real, albeit we do delve deeper with complex functions in our upcoming work \cite{paper}.
%As the ambiguity functions appear in the denominator, after quantization $\pi_{a}$ dependence leads to an inverse differential operator, hence an infinite sum of derivatives. 
%The dimension of the Hilbert space becomes infinite, since the quantum equation becomes of infinite order. Such a candidate of quantum theory cannot be equivalent to that described by a second order differential equation since the Hilbert space of the latter is smaller. So, to have any chance of having a universal theory, we must restrict ourselves to ambiguities that are only functions of $a$. This does not mean that these "new" candidates of quantum theories (with an infinite number of derivatives) are necessarily irrelevant. It could be the case that a second order differential equation is simply an approximation, as higher order derivatives give terms of higher order in $\hbar$, therefore they are considered very small. One would have to examine whether this would lead to contradictions or pathologies. In the absence of such, we cannot in principle disregard a possible $\pi$-dependence for $\rho_{1}$, $\rho_{2}$. \\

Introducing the ambiguity functions, the quantum Hamiltonian gives the Wheeler-DeWitt equation
\begin{equation}\label{wdw}
\begin{split}
\frac{H}{N}\Psi&= -\frac{\hbar ^{2}}{4} \frac{\gamma^{aa}}{\rho_{1}\rho_{2}}\frac{d}{da}\biggl(\rho_{1}\frac{d\left(\rho_{2}\Psi\right)}{da}\biggr)+V\Psi =0.
\end{split}
\end{equation}
The wavefunctions $\Psi$ can be expressed as path integrals \cite{toumbas, dewitt, hartle, halliwell}
\begin{equation} \label{pathintegral}
    \Psi = \int \mathcal{D}a \, e^{\frac{iS[a]}{\hbar}},
\end{equation}
where the path integral measure has to be suitably chosen so as to satisfy (\ref{wdw}) for the given $\rho_1$ and $\rho_2$. 
The states are scalars under field redefinitions (which are expressed here as a change of coordinates), while the ambiguity functions transform as:

\begin{equation}
    \rho_{2}\longrightarrow \rho_{2}\, , \qquad \rho_{1}\longrightarrow \rho_{1} \frac{d\tilde{a}}{da}\equiv \tilde{\rho}_{1} 
\end{equation}

The WDW equation can be brought into a more convenient form
\begin{equation}
\begin{split}
     \frac{H}{N} \Psi &= -\frac{\hbar ^{2}}{4} \gamma^{aa}\left \{\frac{1}{\rho} \frac{d}{da} \left(\rho\, \frac{d\Psi}{da}\right) + \omega \Psi \right\}+V\Psi=0,
\end{split}
\end{equation}
where we have defined:
\begin{equation}
\begin{split}
&\rho = \rho_{1} \rho _{2}^{2} ,~~ \omega = \frac{1}{\rho_{1} \rho _{2}}\frac{d}{da}\left (\rho _{1}\frac{d  \rho_{2}}{da} \right ).
\end{split}
\end{equation}

Next we define the inner product as an integral over the minisuperspace manifold $\mathcal{T}$ as follows
\begin{equation} \label{amp}
    \langle \Psi_{1} \, , \, \Psi_{2} \rangle = \int \limits _{\mathcal{T}} da \, \sqrt{-\gamma } \, \mu(a) \, \Psi_{1} ^{*} \Psi_{2}
\end{equation}
where the real positive function $\mu$ is a suitable measure. 
%It becomes apparent that an arbitrary field redefinition of the scale factor is a diffeomorphism on the minisuperspace manifold. 
Notice that the volume element  $da \, \sqrt{-\gamma}$ remains invariant under reparametrizations of the minisuperspace manifold.
    
We next impose the condition that the Hamiltonian is hermitian. To this end, we consider the following identity involving the hermitian conjugate of the Hamiltonian, obtained by performing an integration by parts
\begin{equation} \label{inner1}
     \Bigg \langle \Psi_{1} \, , \, \frac{H}{N} \Psi_{2} \Bigg \rangle = \Bigg \langle \frac{H^{\dagger}}{N} \Psi_{1} \, , \, \Psi_{2} \Bigg \rangle+\text{Boundary terms}.
\end{equation}
For hermiticity to hold, we must have
\begin{equation} \label{eq:hermiticity}
     H= H^{\dagger},
\end{equation}
while the boundary terms in the identity above must vanish. A proof of the latter can be found in \cite{toumbas, paper}. 

Imposing equation (\ref{eq:hermiticity}) on an arbitrary wavefunction, we obtain the following constraint 
 \begin{equation} \label{eq:constr}
     \frac{d}{da} \left ( \gamma ^{aa} \frac{\mu \sqrt{-\gamma}}{\rho} \right ) = 0,
 \end{equation}
 which fixes the measure to be
 \begin{equation} \label{eq:1dmeasure}
\mu=\kappa \frac{\gamma _{aa} \, \rho}{~\sqrt{-\gamma}}.
 \end{equation}
 Here $\kappa$ is a physically irrelevant constant, which can be set equal to $\pm 1$ accordingly so that the measure is positive-definite. Thus, the measure $\mu$ depends on the ambiguity function $\rho$. 
 %The probability amplitude is by extension also dependent on it. 
 At the semi-classical level, however, it has been demonstrated that the inner products are universal \cite{toumbas}.  In the following we would like to examine and derive conditions on the ambiguity functions for this universality property to hold to all orders in $\hbar$. 
 
For this purpose,  we use the inner product measure to define the wavefunction $\widehat{\Psi}$, which transforms as a scalar under field redefinitions:
% We are going to use a covariant second-order differential operation:
%\begin{equation} \label{eq:nabla}
%    \nabla ^{2} \widehat{\Psi} = \gamma ^{aa} \nabla_{a} \nabla_{a} \widehat{\Psi}
%\end{equation}
%\noindent where $\nabla_{a}$ is the covariant derivative in the minisuperspace metric, and where we have:
\begin{equation} \label{eq:18}
    \widehat{\Psi} = \sqrt{\mu} \, \Psi .
\end{equation}
To cast the WDW equation for $\widehat{\Psi}$ in an explicitly covariant form, we first obtain the invariant D'Alembertian:
\begin{equation} \label{eq:nabla'}
\begin{split}
    &\nabla ^{2} \widehat{\Psi} = \frac{1}{\sqrt{-\gamma}} \left [\sqrt{-\gamma} \, \gamma ^{aa} \left (\sqrt{\mu} \, \Psi \right )' \right ]' = \\
    \sqrt{\mu} \, \gamma ^{aa} \Psi '' + & \gamma ^{aa} \left (\sqrt{\mu} \right )' \, \Psi ' + \frac{1}{\sqrt{-\gamma}} \left (\sqrt{-\gamma} \, \gamma ^{aa} \sqrt{\mu} \right )' \Psi ' + Z \widehat{\Psi},
\end{split}
\end{equation}
where $\nabla$ stands for the covariant derivative with respect to the minisuperspace metric and we have defined
\begin{equation} \label{eq:z}
\begin{split}
    Z=& \frac{1}{\sqrt{\mu} \sqrt{-\gamma}} \left [\sqrt{-\gamma} \, \gamma ^{aa} \left ( \sqrt{\mu} \right )' \right]' \\
    =&-\frac{1}{4}\gamma ^{aa}\left(\frac{\rho^{\prime}}{\rho}\right)^2+ \frac{1}{2}\gamma ^{aa}\frac{\rho^{\prime\prime}}{\rho}-\frac{3}{4}\gamma ^{aa}\left(\frac{(\sqrt{-\gamma})^{\prime}}{\sqrt{-\gamma}}\right)^2 +\frac{1}{2}\gamma ^{aa}\frac{(\sqrt{-\gamma})^{\prime\prime}}{\sqrt{-\gamma}}. 
\end{split}
\end{equation}
The primes indicate ordinary derivatives with respect to the scale factor $a$. Using (\ref{eq:1dmeasure}), we can simmplify the expression of the D'Alembertian:
\begin{equation} \label{eq:nablaexpanded}
    \nabla ^{2} \widehat{\Psi} = \sqrt{\mu} \, \gamma ^{aa}\left ( \Psi '' + \frac{\rho '}{\rho}  \Psi ' \right ) + Z\widehat{\Psi}. 
\end{equation}
Using the WDW equation for $\Psi$, we finally obtain
\begin{equation}\label{eq:wdwco}
 \nabla^2\widehat{\Psi}-\frac{4V}{\hbar^2}\widehat{\Psi}+ \mathcal{Z}\widehat{\Psi}=\sqrt{\mu}\biggl\{\gamma ^{aa}\left \{\frac{1}{\rho} \frac{d}{da} \left(\rho\, \frac{d\Psi}{da}\right) + \omega \Psi \right\}-\frac{4V}{\hbar^2}\Psi\biggr\} = 0.
 \end{equation}
The ambiguity functions appear only in the scalar function
\begin{equation} \label{eq:z'}
    \mathcal{Z} = \gamma ^{aa} \omega - Z,
\end{equation}
which leads to higher order quantum corrections, not present in the semi-classical approximation. The resulting equation is thus free of the ambiguity functions at the semiclassical level, and so are $\widehat{\Psi}$ and the inner products
\begin{equation} \label{eq:amp'}
    \langle \Psi_{1} \, , \, \Psi_{2} \rangle = \int \limits _{\mathcal{T}} da \, \sqrt{-\gamma } \, \widehat{\Psi}_{1} ^{*} \widehat{\Psi}_{2}.
\end{equation}

In order for the universality to extent to all orders, it is necessary that $\mathcal{Z}$ is some universal scalar quantity, independent of the ambiguity functions. This fixes the ambiguity function $\omega$ in terms of $\rho$. The no boundary path integral wavefunctions of \cite{toumbas} may belong to the same universality class in that they share the same $\mathcal{Z}$, irrespective of $\rho$. As anticipated in \cite{toumbas}, $\omega$ is determined in terms of $\rho$ for these wavefunctions and the relation could be such that $\mathcal{Z}$ is ambiguity free. This is investigated in a forthcoming publication \cite{paper}. 

%we can bring the WDW equation to a Schrödinger form which would necessitate $\omega = 0$ and subsequently $\mathcal{Z} = 0$ \cite{toumbas, paper}. This however does not extend to all other choices of ambiguity functions, since there is no empirical data to suggest that such a quantum system would behave like a point particle under the influence of a potential $V$. Hence the best we can do is assert that the inner product (\ref{eq:amp'}) is universal to all orders in order for the observables to remain the same for every choice of ambiguity. 

\section{Minisuperspace of arbitrary dimensions}

\noindent We consider a gravitational action with a positive cosmological constant $\Lambda$ and include homogeneous scalar matter fields, which we take to be minimally coupled to gravity:
\begin{equation} \label{eq:0}
    S = \frac{1}{2} \int \limits_{\mathcal{V}} d^{d +1}x \, \sqrt{-g} \, \left [R-2\Lambda - g^{00} \sum _{\alpha ,\beta} C_{\alpha \beta} \dot{\phi}^{\alpha} \dot{\phi}^{\beta} - 2V \left (\left \{\phi ^{\alpha} \right \} \right ) \right ].
\end{equation}
Retaining homogeneity, the spacetime metric can be written in the form \cite{dewitt, hawking, wheeler}:
\begin{equation} \label{eq:2}
    ds^{2} = -\left [N^{2}(t) - N_{i}(t)N^{i}(t) \right] dt^{2} +2N_{i}(t)dt \, dx^{i} + h_{ij}(t) dx^{i} dx^{j} 
\end{equation}
where $N$ is the lapse function, $N_{i}$ are shift factors and $h_{ij}$ stand for the spatial components of the metric. The scalar fields interact via the potential term in the Lagrangian. $C_{\alpha \beta} = C_{\beta \alpha}$ is the metric on the scalar field manifold, which we take to be non-trivial. 
%These can be understood as elements of a $\phi$-space metric, and they are generally taken to be dependent on all the fields ($h_{ij}$, $\phi ^{\alpha}$). They are not allowed to be dependent on $\dot{h}_{ij}$ or $\dot{\phi}^{\alpha}$. Allowing the former would translate into a non-minimal coupling between matter and gravity, since one may express $\dot{h}_{ij}$ as a function of the Ricci scalar $R$. Allowing the latter would render the term non-kinetic, as it would cease to be quadratic in $\dot{\phi}^{\alpha}$.\\

As before, we will be working in the minisuperspace approximation, taking the Universe to be closed and homogeneous. This time, however, we allow for anisotropies. As a result, the spatial metric $h_{ij}$ will give rise to $D={d(d+1)}/{2}$ independent degrees of freedom. A suitable combination of these variables corresponds to the dilatation of the spatial manifold, while the rest describe other geometrical properties related to the shape of the spatial manifold \cite{dewitt, gibbons2, demianski}. The independent dynamical metric degrees of freedom are taken to be
\begin{equation} \label{eq:23}
\begin{split}
    &a(t) = f_{0} \left ( \{ h_{ij} \} \right ) \\
    &\eta^{k}(t) = f^{k} \left ( \{ h_{ij} \} \right ) \quad , \, k \in \{1,\dots, D-1\}.
\end{split}
\end{equation}
%The forms of these functions are such that they are independent between them.  When reformulating the Lagrangian in these terms, we will find that 
Of all these only $a$ has a negative kinetic term; the remaining $D-1$ degrees of freedom have positive kinetic terms \cite{demianski}. 
Taking into account the $m$ scalar matter fields $\phi ^{\alpha}$, we end up with $M=m+D-1$ fields with positive kinetic terms. 

The full set of fields in the model is thus described as follows:
\begin{equation} \label{eq:3}
    q^{A} =
\begin{cases}
    &a \, , \quad A = 0 \\
    &\eta^{k} \, , \quad A = k \in \left \{1,\dots, D-1 \right \} \\
    &\phi^{\alpha} \, , \quad A = \alpha + D-1 \in \left \{D,\dots, M \right\}.
\end{cases}
\end{equation}
The lapse function and shift factors $N$ and $N_{i}$ have no kinetic terms, as they not dynamical variables. 

Upon carrying out the spatial integrals in the action (\ref{eq:0}), we obtain the effective potential of the theory given by
\begin{equation} \label{eq:V}
    \widetilde{V} \left ( \left \{ q^{A} \right \} \right ) = \frac{1}{2} \int \limits _{\partial \mathcal{V}} d^{d}x \sqrt{-g} \, \left [ 2V \left (\left \{\phi ^{\alpha} \right \} \right ) + 2\Lambda - {}^{d}R\right ],
\end{equation}
where ${}^{d}R$ is the curvature on a constant time slice $\partial \mathcal{M}$ of metric $h_{ij}$. In terms of this, we can cast the theory as a non-linear $\sigma$-model with a one-dimensional base manifold and a Lorentzian target space parametrized by the dynamical fields $q^{A}$: 
\begin{equation} \label{eq:1}
    L = N\left[\gamma _{A B} \,  \dot{q}^{A} \dot{q}^{B} - \widetilde{V} \left (\left \{ q^{A} \right \} \right ) \right ],
\end{equation}
where $\gamma _{A B}$ is the minisuperspace metric, or the metric on the target space manifold. This metric is of Lorentzian signature since $q^{0}$ is time-like. Notice that $\gamma _{AB}$ turns out to be block-diagonal - one block corresponding to the gravitational components and the other to the matter components. This is because there is no kinetic term in the Lagrangian combining $\dot{h}_{ij}$ and $\dot{\phi} ^{\alpha}$. However, under generic field redefinitions of $q^{A}$, there could be mixings. In such a frame, the minisuperspace metric will no longer retain its block diagonal form.

From the Lagrangian (\ref{eq:1}), it is straightforward to derive the conjugate momenta $\pi_A$ and the classical Hamiltonian, which takes the form
\begin{equation} \label{eq:4}
    \frac{H}{N} = \frac{1}{4} \gamma ^{A B} \pi _{A} \pi _{B} + \widetilde{V} \left ( \left \{ q^{A} \right \} \right )
\end{equation}
In order to quantize the Hamiltonian, we promote $q^A$ and $\pi_A$ to operators and impose canonical commutation relations:
\begin{equation} \label{eq:5}
\begin{split}
    \begin{split}
    [q^{A},\, \pi _{B}] &= i\hbar \, \delta ^{A}{} _{B} \\
     [q^{A}, q^{B}] &= 0 \\
    [\pi _{A} , \pi _{B}] &= 0.
\end{split}
\end{split}
\end{equation}
Therefore, when acting on scalar wavefuctions the conjugate momenta of the classical theory are promoted to partial derivatives:
\begin{equation} \label{eq:6}
    \pi _{A} \rightarrow -i\hbar \frac{\partial}{\partial q^{A}}.
\end{equation}

Just like in the $1d$ case, the quantum Hamiltonian is not uniquely defined, but there are operator ordering ambiguities. Analogously to (\ref{eq:ambiguities}), we introduce sets of ambiguity functions capturing a large class of models\footnote{
A more general treatment would have been to introduce ambiguity functions as matrices in the form $\gamma ^{AB} \rho^{EF}_{(2)} \rho^{CD}_{(1)} \pi _{C} \, \rho^{(1)}_{AD} \pi_{E} \, \rho^{(2)}_{BF}$. This would allow some extra quantum terms not appearing in the current paper. A more detailed analysis is described in our forthcoming paper \cite{paper}.}:
\begin{equation} \label{eq:7}
    \gamma ^{A B} \pi _{A} \pi _{B} = \frac{1}{\rho _{1}^{(A;B)} \rho _{2}^{(A;B)}} \gamma ^{A B} \pi_{A} \, \rho _{2}^{(A;B)} \, \pi _{B} \, \rho _{1}^{(A;B)}.
\end{equation}
Each of these functions depends generically on all fields $q^{A}$. We emphasize that the ordering of the indices in the superscripts of the $\rho$'s  plays a role. In other words,  $\rho^{(A;B)}\neq \rho^{(B;A)}$ unless $A = B$.  We end up with $2(M+1)^{2}$ ambiguity functions, which may appear in the expression of the quantum Hamiltonian. It is important to mention that just like in section 2, we restrict ourselves to real ambiguity functions, leaving the complex case for a later paper \cite{paper}.

It will be useful to define the following functions
\begin{equation} \label{eq:8}
\begin{split}
    &\rho_{(A;B)} = \rho _{1}^{(A;B)} \rho _{2}^{(A;B)} \rho _{2}^{(B;A)} \\
    &\omega_{(A;B)} = \frac{\partial _{A} \left [\rho _{1}^{(A;B)} \partial _{B} \left (\rho _{2}^{(A;B)} \right ) \right ]}{\rho _{1}^{(A;B)} \, \rho _{2}^{(A;B)}},
\end{split}
\end{equation}
which will appear in the WDW equation for the wavefunction of the Universe. There are $(M+1)^{2}$ $\rho$ functions and $(M+1)^{2}$ independent $\omega$ functions. For further convenience, we define the symmetric combination
\begin{equation}
    \Omega_{(A;B)} = \frac{1}{2} \gamma ^{A B} \left (\omega_{(A;B)} + \omega_{(B;A)} \right ).
\end{equation}
Since $\Omega_{(A;B)} = \Omega_{(B;A)}$, there are only $\left ({M}/{2} +1 \right )(M+1)$ independent such functions. 

As before, the quantum Hamiltonian must annihilate the wavefunction of the Universe $\Psi$, which is a function of the $M+1$ $q^A$, resulting in the WDW equation
\begin{equation} \label{eq:9}
    \frac{H}{N} \Psi(\{q^A\}) = 0.
\end{equation}
In terms of the ambiguity functions the WDW equation is given by
\begin{equation} \label{eq:10}
    -\frac{\hbar ^{2}}{4} \sum _{A,B = 0}^{M} \left [\gamma ^{A B} \frac{1}{\rho_{(A;B)}} \frac{\partial}{\partial q^{A}} \left( \rho_{(A;B)} \frac{\partial \Psi}{\partial q^{B}} \right ) + \Omega_{(A;B)} \Psi \right ] + \widetilde{V} \Psi = 0.
\end{equation}
The sums over $A$ and $B$ must be explicitly written to avoid confusion with the usual convention for contractions.  The solutions to this equation are in general dependent on the ambiguity functions. 

The inner product is defined as an integral over the minisuperspace manifold $\mathcal{T}$:
\begin{equation} \label{eq:11}
    \langle \Psi_{1} \, , \, \Psi_{2} \rangle = \int \limits_{\mathcal{T}} d^{1+M} q \, \sqrt{-\gamma } \, \mu(q^{A}) \, \Psi_{1} ^{*} \Psi_{2}
\end{equation}
where $\mu$ is a suitable measure. We will now proceed to investigate whether these inner products are universal, free of any of the ambiguity functions. 

As in the 1d case, we first impose hermiticity for the quantum Hamiltonian. The conjugate of the Hamiltonian is the operator satisfying
\begin{equation} \label{eq:12}
     \Bigg \langle \Psi_{1} \, , \, \frac{H}{N} \Psi_{2} \Bigg \rangle = \Bigg \langle \frac{H^{\dagger}}{N} \Psi_{1} \, , \, \Psi_{2} \Bigg \rangle + {\rm boundary\, terms}.
\end{equation}
It can be shown explicitly that the boundary terms in the above identity vanish. Hence, to establish hermiticity we must have 
\begin{equation} \label{eq:13}
    H = H^{\dagger}.
\end{equation}
Using integration by parts and (\ref{eq:12}), we obtain 
\begin{equation} \label{eq:14}
    \frac{H^{\dagger}}{N} \Psi = -\frac{\hbar ^{2}}{4} \sum _{A,B = 0}^{M} \left \{\frac{1}{\mu \sqrt{-\gamma }} \frac{\partial}{\partial q^{B}} \left [\rho_{(A;B)} \frac{\partial}{\partial q^{A}}  \left (\gamma ^{A B}\frac{\mu \sqrt{-\gamma }}{\rho_{(A;B)}} \Psi \right ) \right] + \Omega_{(A;B)} \Psi \right \} + \widetilde{V} \Psi 
\end{equation}
and demand that (\ref{eq:13}) holds. All terms besides the ones involving the $\rho$ functions are trivially the same as the corresponding ones in the expression for $H$. Hence these cancel trivially between the two sides of (\ref{eq:13}). Likewise the second derivative terms cancel. So we end up with the following constraint:
\begin{equation} \label{eq:14'}
\begin{split}
    &\sum _{A,B = 0}^{M} \gamma ^{A B} \frac{1}{\rho_{(A;B)}} \frac{\partial \rho_{(A;B)}}{\partial q^{A}} \frac{\partial \Psi}{\partial q^{B}} = \frac{1}{\mu \sqrt{-\gamma }} \sum _{A,B = 0}^{M} \Bigg \{  \frac{\partial \rho_{(A;B)}}{\partial q^{B}} \frac{\partial}{\partial q^{A}}  \left (\gamma ^{A B}\frac{\mu \sqrt{-\gamma }}{\rho_{(A;B)}} \Psi \right ) \\
    + &\rho_{(A;B)} \frac{\partial \Psi}{\partial q^{A}} \frac{\partial}{\partial q^{B}}  \left (\gamma ^{A B}\frac{\mu \sqrt{-\gamma }}{\rho_{(A;B)}}\right ) + \rho_{(A;B)} \frac{\partial \Psi}{\partial q^{B}} \frac{\partial}{\partial q^{A}}  \left (\gamma ^{A B}\frac{\mu \sqrt{-\gamma }}{\rho_{(A;B)}}\right ) + \rho_{(A;B)} \frac{\partial ^{2}}{\partial q^{B} \partial q^{A} }  \left (\gamma ^{A B}\frac{\mu \sqrt{-\gamma }}{\rho_{(A;B)}} \right )\Psi \Bigg \}.
\end{split}
\end{equation}

In this expression, there are terms linear in $\Psi$ and terms linear in a partial derivative of $\Psi$. Since $\Psi$ is an arbitrary solution of the WDW equation, the terms linear in $\Psi$ and the terms linear in the derivatives of $\Psi$ must vanish separately. Therefore, we obtain the following two equations:
\begin{equation} \label{eq:constraints1}
\begin{split}
    &\sum _{A,B = 0}^{M} \gamma ^{A B} \frac{1}{\rho_{(A;B)}} \frac{\partial \rho_{(A;B)}}{\partial q^{A}} \frac{\partial \Psi}{\partial q^{B}} = \sum _{A,B = 0}^{M} \Bigg \{ \gamma^{AB} \frac{1}{\rho_{(A;B)}} \frac{\partial \rho_{(A;B)}}{\partial q^{B}} \frac{\partial \Psi}{\partial q^{A}} \\
    + &\frac{\rho_{(A;B)}}{\mu \sqrt{-\gamma }} \left [\frac{\partial \Psi}{\partial q^{A}} \frac{\partial}{\partial q^{B}}  \left (\gamma ^{A B}\frac{\mu \sqrt{-\gamma }}{\rho_{(A;B)}}\right ) + \frac{\partial \Psi}{\partial q^{B}} \frac{\partial}{\partial q^{A}}  \left (\gamma ^{A B}\frac{\mu \sqrt{-\gamma }}{\rho_{(A;B)}}\right ) \right ]\Bigg \}
\end{split}
\end{equation}
and
\begin{equation} \label{eq:constraints2}
    \frac{1}{\mu \sqrt{-\gamma }} \sum _{A,B = 0}^{M} \Bigg \{ \frac{\partial \rho_{(A;B)}}{\partial q^{B}} \frac{\partial}{\partial q^{A}}  \left (\gamma ^{A B}\frac{\mu \sqrt{-\gamma }}{\rho_{(A;B)}} \right ) + \rho_{(A;B)} \frac{\partial ^{2}}{\partial q^{B} \partial q^{A} }  \left (\gamma ^{A B}\frac{\mu \sqrt{-\gamma }}{\rho_{(A;B)}} \right ) \Bigg \}\Psi = 0.
\end{equation}

By expanding some of the derivative terms, the first equation simplifies to
%\begin{equation} \label{eq:constraints3}
%\begin{split}
 %   \sum _{A,B = 0}^{M} \gamma ^{A B} \frac{1}{\rho_{(A;B)}} \frac{\partial \rho_{(A;B)}}{\partial q^{A}} \frac{\partial \Psi}{\partial q^{B}} = \sum _{A,B = 0}^{M} %\frac{1}{\mu \sqrt{-\gamma}} \frac{\partial}{\partial q^{B}} \left (\gamma ^{AB} \mu \sqrt{-\gamma} \right ) \frac{\partial \Psi}{\partial q^{A}} \\
 %   + \frac{1}{\mu \sqrt{-\gamma}} \frac{\partial}{\partial q^{A}} \left (\gamma ^{AB} \mu \sqrt{-\gamma} \right ) \frac{\partial \Psi}{\partial q^{B}} - \sum _{A,B = %0}^{M} \gamma ^{AB} \frac{1}{\rho_{(A;B)}} \frac{\partial \rho_{(A;B)}}{\partial q^{A}} \frac{\partial \Psi}{\partial q^{B}} 
%\end{split}
%\end{equation}
%\noindent The first and second terms in the rhs are symmetric in the indices $A,B$ due to all of their possible combinations occurring in the double sum, %and are therefore equal. The last term in the rhs is identically the same as the one on the lhs. As a result we have:
\begin{equation} \label{constraints4}
    2\left \{\sum _{A,B = 0}^{M} \gamma ^{A B} \frac{1}{\rho_{(A;B)}} \frac{\partial \rho_{(A;B)}}{\partial q^{A}}  - \frac{1}{\mu \sqrt{-\gamma}} \frac{\partial}{\partial q^{A}} \left (\gamma ^{AB} \mu \sqrt{-\gamma} \right ) \right \} \frac{\partial \Psi}{\partial q^{B}} = 0.
\end{equation}
For the equation above to hold for arbitrary $\Psi$, the coefficient of each partial derivative $\partial _{B} \Psi$ should vanish separately. 
%Rearranging the terms and putting them in a more compact form, 
We end up with a set of $M+1$ constraints:
\begin{equation} \label{eq:15}
    \sum _{A = 0} ^{M} \frac{\partial}{\partial q^{A}} \left (\gamma ^{A B} \frac{\mu \sqrt{-\gamma }}{\rho_{(A;B)}} \right ) \rho_{(A;B)} = 0.
\end{equation}
Applying these constraints to the lhs of equation (\ref{eq:constraints2}), we find that the equality indeed holds.
%meaning the terms linear in $\Psi$ also vanish. 
As a result, (\ref{eq:13}) is satisfied. Notice that in the higher dimensional case the measure $\mu$ is not unique, but it is constrained by (\ref{eq:15}).

The steps to find classes of models with universal observables are similar to those in the one-dimensional case. We define the scalar function 
$\widehat{\Psi} = \sqrt{\mu} \, \Psi$  as in (\ref{eq:18}) and obtain the WDW equation it satisfies.
The D'Alembertian operator acting on $\widehat{\Psi}$  yields:
\begin{equation} \label{eq:step1}
\begin{split}
    &\nabla ^{2} \widehat{\Psi} = \frac{1}{\sqrt{-\gamma}} \partial _{A} \left [\sqrt{-\gamma} \, \gamma ^{AB} \partial _{B} \left (\sqrt{{\mu}} \, \Psi \right ) \right ] = \\
    \sqrt{{\mu}} \, \gamma ^{AB} \partial _{A} \partial _{B} \Psi + & \gamma ^{AB} \left (\partial _{B} \sqrt{{\mu}} \right ) \, \partial _{A} \Psi + \frac{1}{\sqrt{-\gamma}} \partial _{A} \left (\sqrt{-\gamma} \, \gamma ^{AB} \sqrt{{\mu}} \right ) \partial _{B} \Psi + Z \widehat{\Psi},
\end{split}
\end{equation}
where $Z$ is given by
\begin{equation} \label{eq:18'}
    Z = \frac{1}{ \sqrt{-\mu\,\gamma} } \partial _{A} \left [ \sqrt{-\gamma} \, \gamma ^{AB} \left ( \partial _{B} \sqrt{{\mu}} \right ) \right ]
\end{equation}
generalizing (\ref{eq:z}). Using (\ref{eq:15}) and some algebra, we finally obtain
\begin{equation} \label{eq:step5}
    \nabla ^{2} \widehat{\Psi} = \sqrt{{\mu}}\sum _{A,B} \gamma ^{AB} \left (\frac{\partial ^{2} \Psi}{\partial q^{A} \partial q^{B}} +  \frac{1}{{\rho} _{(A;B)}} \frac{\partial {\rho}_{(A;B)}}{\partial q^{A}}  \frac{\partial \Psi}{\partial q^{B}} \right ) + Z\widehat{\Psi}.
\end{equation}
Noticing that the sum is none other than the derivative part of the WDW equation Eq.~(\ref{eq:10}), we obtain:
\begin{align} 
  \sqrt{ \mu}  \Bigg \{ \sum _{A,B}^M \Bigg [\gamma ^{A B} \frac{1}{{\rho}_{(A;B)}} \frac{\partial}{\partial q^{A}} \left( {\rho}_{(A;B)} \frac{\partial \Psi}{\partial q^{B}} \right ) &+ \Omega_{(A;B)} \Psi \Bigg ] -\frac{4 \color{black}\widetilde{V}}{\color{black}\hbar^2} \Psi \Bigg \} = \\
  \nabla ^{2} \widehat{\Psi} + \left (\mathcal{Z}-\frac{4\widetilde V}{\hbar^2} \right )\widehat\Psi = 0,
   \label{eq:17'}
\end{align}
where
\begin{equation} \label{eq:00}
    \mathcal{Z} = -Z + \sum _{A,B}^M \Omega_{(A;B)},
\end{equation}
generalizing the equation satisfied by $\widehat{\Psi}$ in the one dimensional case. 

Some remarks are in order:
\begin{itemize}
\item At the semiclassical level, the WDW equation of $\widehat{\Psi}$ is universal, independent of the ambiguity functions. From Eq.~(\ref{eq:17'}) we see that the dependence on the ambiguity functions arises from $\mathcal{Z}$, but this term leads to higher order corrections in $\hbar$, which do not contribute at the semiclassical level. Since the inner product and all probability densities can be determined in terms of $\widehat{\Psi}$ and the invariant volume element in minisuperspace, we conclude that the predictions are universal at the semiclassical level, independent of any operator ordering ambiguities.
%(\ref{eq:11}) which we rewrite as:
%\begin{equation} \label{eq:20}
 %   \langle \Psi_{1} \, , \, \Psi_{2} \rangle = \int \limits_{\mathcal{M}} d^{1+M} q \, \sqrt{-\gamma } \, \widehat{\Psi}_{1} ^{*} \widehat{\Psi}_{2}
%\end{equation}
%\noindent The volume element of the manifold as well as the scalars $\widehat{\Psi}_{i}$ are independent of the ambiguity functions and thus universal for any choice of operator ordering in the WDW Hamiltonian. 
\item Beyond the semi-classical level, there is in general dependence on the ordering ambiguities via $\mathcal{Z}$. Notice that this must transform as a scalar under field reparametrizations, as the effective potential $\widetilde{V}$. We can construct classes of models with universal predictions within the class if we demand that $\mathcal{Z}$ is free of ambiguities. This imposes special relations among the ambiguity functions. One class of models are those for which $\mathcal{Z}$ vanishes. However, $\mathcal{Z}$ can be any scalar quantity, invariant under reparametrizations of the minisuperspace manifold.  
\item As in the one dimensional case, it would be interesting to see if the path integral no boundary wavefunctions fall in such categories for which $\mathcal{Z}$ is ambiguity free. In particular, we would like to check if they fall in the same class, and therefore yielding the same predictions. The universality at the semiclassical level would then extend to all orders in $\hbar$. 
\end{itemize}

\section{Conclusions}

\noindent The formalism we present in this paper has primarily made use of the hermiticity of the Hamiltonian and the invariance of the theory under field redefinitions. It has proved adequate in confirming the universality of the 1d minisuperspace inner product at the semi-classical level. We also argued that this universality property at the semiclassical level can be generalized to minisuperspace models of arbitrary dimensions. It has proven very useful to define the scalar wavefunction $\widehat{\Psi}$ in terms of the inner product measure that renders the Hamiltonian hermitian and obtain the WDW equation it satisfies. This equation is invariant under field redefinitions of the minisuperspace variables and all operator ordering ambiguity functions appear in higher order terms in $\hbar$, a property that allows one to construct classes of models with universal predictions to all orders. In subsequent work \cite{paper}, we show that the path integral wavefunctions of \cite{toumbas} form such a universality class, at least for the 1d minisuperspace case.

\section*{Acknowledgements}

\noindent We would like to thank Herv\'e Partouche and Nicolaos Toumbas for useful discussions and guidance. This work is partially supported by the Cyprus Research and Innovation Foundation grant EXCELLENCE/0421/0362 and by the European Regional Development Fund.

\end{document}